\documentclass[
    aip,
    amsmath,
    amssymb,
    preprint, 
    superscriptaddress
]{revtex4-2}

\usepackage{graphicx} 
\usepackage{bm}
\usepackage{siunitx}
\DeclareSIUnit{\torr}{Torr}
\usepackage[version=4]{mhchem}
\usepackage{booktabs}
\usepackage{hyperref}
\usepackage{comment}

\graphicspath{{figures/}}

\begin{document}

\title{
Growth and Interface Engineering of Superconducting TiN on Sapphire by Thermal-Laser Epitaxy
}

\author{Anthony Hyatt}
\affiliation{
Department of Materials Science and Engineering,
Cornell University,
Ithaca, New York 14853, USA
}

\author{Anand Ithepalli}
\affiliation{
Department of Materials Science and Engineering,
Cornell University,
Ithaca, New York 14853, USA
}

\author{Eegene Clara Chung}
\affiliation{
Department of Physics,
Cornell University,
Ithaca, New York 14853, USA
}

\author{Yorick A. Birkh\"olzer}
\affiliation{
Department of Materials Science and Engineering,
Cornell University,
Ithaca, New York 14853, USA
}

\author{Brendan Faeth}
\affiliation{
Department of Materials Science and Engineering,
Cornell University,
Ithaca, New York 14853, USA
}
\affiliation{epiray United States, Springfield, Virginia 22152, USA}

\author{Huili Grace Xing}
\affiliation{
Department of Materials Science and Engineering,
Cornell University,
Ithaca, New York 14853, USA
}

\author{David A. Muller}
\affiliation{School of Applied and Engineering Physics, Cornell University, Ithaca, NY, USA}
\affiliation{Kavli Institute at Cornell for Nanoscale Science, Ithaca, New York 14853, USA}

\author{Darrell G. Schlom}
\affiliation{
Department of Materials Science and Engineering,
Cornell University,
Ithaca, New York 14853, USA
}
\affiliation{Kavli Institute at Cornell for Nanoscale Science, Ithaca, New York 14853, USA}
\affiliation{Leibniz-Institut f\"ur Kristallz\"uchtung, Max-Born-Str. 2, 12489 Berlin, Germany}

\author{Debdeep Jena}
\email{djena@cornell.edu}
\affiliation{
Department of Materials Science and Engineering,
Cornell University,
Ithaca, New York 14853, USA
}

\date{\today}

\begin{abstract}
Thermal-laser epitaxy (TLE) extends the accessible pressure, temperature, and growth-rate regimes of conventional molecular-beam epitaxy, enabling \textit{in situ} laser annealing, high-purity buffer-layer growth, and efficient evaporation of refractory elements. These capabilities make TLE a promising platform for engineering low-loss superconducting resonators and Josephson-junction heterostructures. Here, we study the TLE growth of TiN on sapphire and observe improved transport properties with increasing growth temperature up to \SI{1150}{\degreeCelsius}. Unfortunately, we observe that the high ammonia pressure and elevated substrate temperatures required for optimal TiN properties promote reactions at the sapphire surface, resulting in voids at the TiN–substrate interface. These defects increase interfacial surface area, introduce dangling bonds, and could compromise tunnel-barrier heterostructures. We mitigate this degradation using an initial TiN seed layer grown at \SI{850}{\degreeCelsius}. With the seed layer, we achieve a superconducting transition temperature of \SI{5.8}{\kelvin}, a residual resistivity ratio of 12.2, and a resistivity of 1.19 \textmu$\Omega\cdot$cm at 10~K, which, to our knowledge, is the lowest reported for TiN grown on sapphire.
\end{abstract}

\maketitle

\section{Introduction and Motivation}
\label{sec:introduction}


Superconducting quantum processors have experienced rapid improvements in coherence over the past decade, driven largely by advances in device architecture and fabrication while continuing to rely predominantly on Al/AlO$_x$/Al Josephson junctions. As coherence times approach the millisecond regime, materials-related loss mechanisms have become an increasingly important limitation. In particular, microwave loss is frequently attributed to two-level systems (TLS), microscopic defects that can couple resonantly to the electromagnetic fields of superconducting circuits.\cite{Martinis2005} Although the microscopic origins of TLS remain an active area of research, proposed sources include amorphous oxides, dangling bonds, chemical disorder, lattice defects, and contamination at surfaces and interfaces.\cite{Muller2019TLS} The contribution of these defects to dielectric loss depends not only on their density, but also on their participation. Participation is defined as the fraction of the device electric-field energy stored in the defective material.\cite{Wang2015Participation} Reducing both defect density and participation therefore remains a major materials challenge for improving superconducting quantum devices.


Molecular-beam epitaxy (MBE) has a long history of addressing materials and interface limitations in semiconductor devices, most notably through the growth of modulation-doped GaAs/AlGaAs heterostructures with exceptionally high electron mobilities.\cite{Dingle1978,Pfeiffer1989} Bringing this degree of materials purity and interface control to superconducting quantum circuits provides a promising route toward reducing materials-related decoherence, as demonstrated by an approximately 80\% reduction in TLS-associated spectral splittings using epitaxial Al$_2$O$_3$ tunnel barriers.\cite{Oh2006} Thermal-laser epitaxy (TLE) extends this approach by applying similar physical principles as conventional MBE while expanding the accessible pressure, temperature, and growth-rate regimes. Introduced in its modern form in 2019,\cite{Groh_1968,Hass1969CO2Laser,Braun2019TLE} TLE uses continuous-wave lasers to heat the source materials and substrate. This enables reactive gas pressures up to $(0.5$–$1.5)\times10^{-3}$ Torr during growth, at which the mean free path of source species equals the source-to-substrate distance.\cite{SCHLOM1995505,Lang_Speck_2012} Even higher pressures are accessible during diffusive growth or \textit{in situ} annealing. TLE also enables substrate temperatures above 2000~$^\circ$C and epitaxial growth rates approaching 1~\textmu m/h.\cite{Smart2021TLE,Majer2024Sapphire} In particular, TLE excels where material synthesis requires high pressures of reactive gases that would harm the filaments in conventional MBE effusion cells.

These capabilities of TLE are particularly attractive for engineering the surfaces and interfaces of superconducting devices. Substrate temperatures exceeding 1700~$^\circ$C enable \textit{in-situ} reconstruction of polished sapphire surfaces,\cite{Sapphire_annealing,Smart_TiN_resonator} while elevated source temperatures and reactive gas pressures enable rapid epitaxial growth, including high-purity sapphire at rates approaching 2~\textmu m/h.\cite{Majer2024Sapphire} Aggressive substrate preparation with thick crystalline buffer and capping layers offers largely unexplored strategies for loss engineering in superconducting circuits. Performing these processes without breaking vacuum may further reduce interfacial defects and their participation in microwave loss. Realizing these advantages requires establishing superconducting thin-film growth under the unusually high temperatures and reactive gas pressures accessible by TLE. This work addresses that prerequisite by investigating and optimizing the growth of TiN within this regime.

Titanium nitride is an established superconducting material for microwave resonators, with a critical temperature near 6~K and reported low-photon internal quality factors often exceeding $10^6$.\cite{Terai2026NitrideQubits,Smart_TiN_resonator,Trenched_TiN,sputtered_RT,Ithepalli2026HighQ,olson2015growth} TiN(111) forms a favorable near coincident site lattice\cite{Balluffi_1982} with Al$_2$O$_3$(0001), enabling epitaxial growth on sapphire, a substrate with exceptionally low microwave loss.\cite{Read2023SapphireLoss} Unfortunately, TiN growth on sapphire is complicated by the formation of double-positioning twin domains and a columnar microstructure, both of which can introduce extended defects despite the favorable epitaxial relationship.\cite{Gao2022TiN,TLE_Nitrides} In contrast to trigonal sapphire, TiN shares the cubic rocksalt structure of MgO with a lattice mismatch of approximately 0.5\% between TiN and MgO.\cite{NTT_TiN_on_MgO} Combined with recent advances in high-purity MgO crystals,\cite{5N_MgO} this compatibility provides a potential pathway toward fully epitaxial TiN/MgO/TiN heterostructures in which the superconducting electrodes and tunnel barrier are crystalline. TiN therefore provides both a technologically relevant superconducting material for establishing TLE growth and a building block for future all-crystalline quantum-device heterostructures.


In this work, we explore TiN growth on sapphire in a regime where both the substrate temperature and reactive nitrogen pressure exceed those conventionally used. These conditions may improve adatom mobility and promote high-quality TiN growth, but they also introduce the possibility of reactions between the growth environment and the substrate. These competing effects expose a tradeoff in TLE growth: high temperatures and reactive nitrogen pressures which favor high-quality TiN can simultaneously destabilize the substrate interface. This is particularly problematic for superconducting devices, where increased interfacial area, chemical disorder, and structural defects may introduce additional microwave loss. This prompts research on the temperature regime where sapphire remains stable under ammonia exposure. These findings are used to develop a low-temperature TiN nucleation layer which lessens substrate degradation during subsequent high-temperature growth.

\section{Experimental Methods}
\label{sec:methods}

\begin{figure}[htbp]
    \centering
    \includegraphics[width=0.5\linewidth]{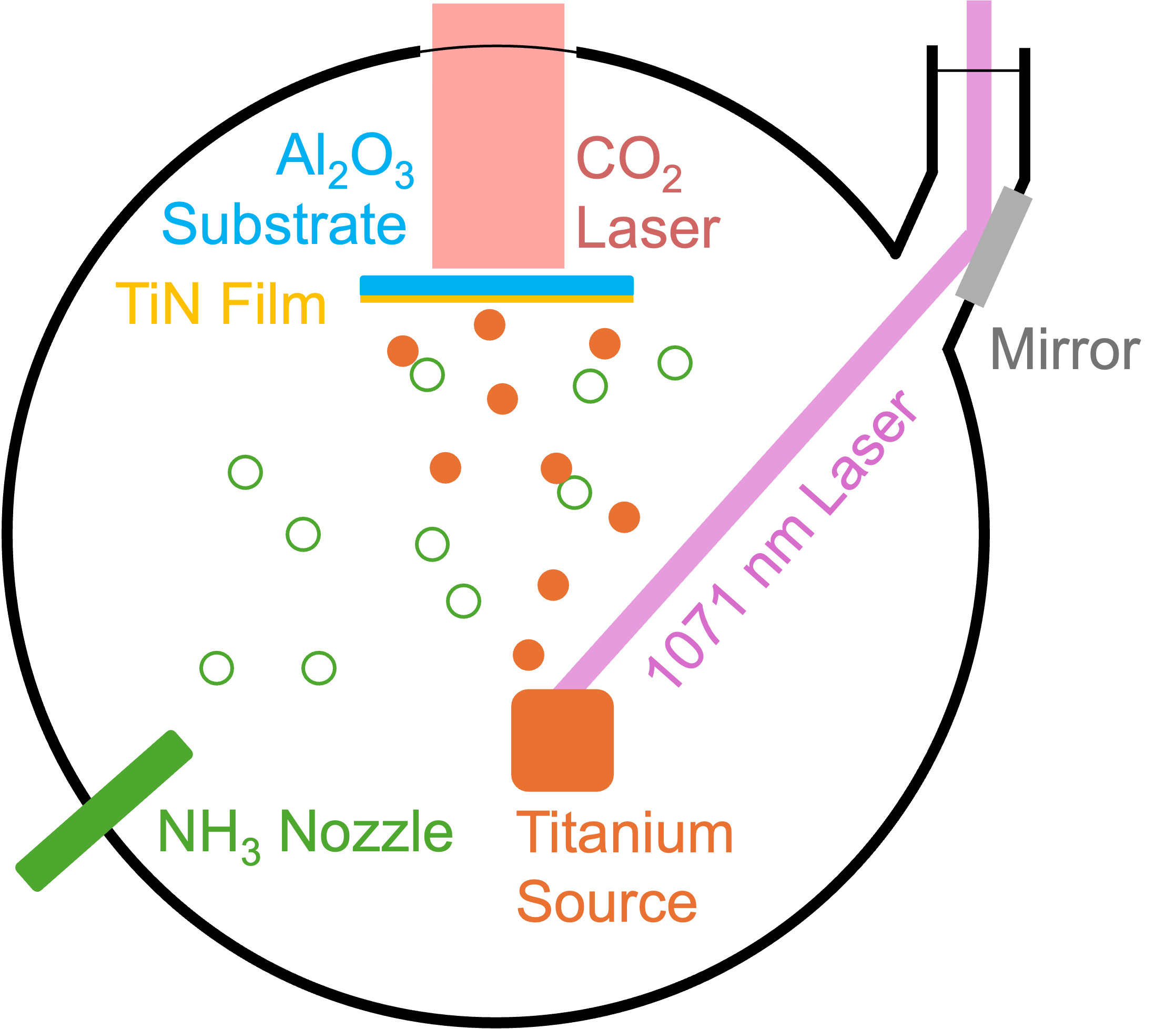}
    \caption{
        Schematic of the TLE configuration used for TiN growth on sapphire, showing laser heating of the substrate and elemental titanium source under an NH$_3$ atmosphere.
    }
    \label{fig:chamber}
\end{figure}

\subsection{Thermal-Laser Epitaxy of TiN}

TiN films were grown using an Epiray STRATOLAS TLE system, schematically illustrated in Fig.~\ref{fig:chamber}, on single-side polished 10~mm $\times$ 10~mm (0001) sapphire substrates purchased from Kyocera, Japan. Substrates were heated from the backside using a 1~kW 10.6~\textmu m CO$_2$ laser (Coherent Diamond J-1000). The substrate holder was machined from corrosion-resistant Haynes 214 superalloy. Titanium was supplied from a 99.995\% pure elemental rod approximately 8 mm tall and 13 mm in diameter, heated by a continuous-wave 1071~nm laser (Trumpf TruFiber Compact P, 500~W) with a 5~mm spot diameter. The titanium source rod was held in place by thin tantalum foil holders at a working distance of 75~mm from the substrate. Reactive nitrogen species were supplied using NH$_3$ gas at a chamber pressure of $1\times10^{-3}$~Torr. The growth chamber had a base pressure of approximately $5\times10^{-9}$ Torr, evacuated using turbomolecular pumps. Substrate temperature was monitored from the backside using a mid-infrared pyrometer (Heitronics KT15.99) centered at 7.5~\textmu m. The calibration of this setup to the melting point of sapphire was previously described by Birkh\"olzer \textit{et~al.}\cite{Birkholzer2026TaO2}

Prior to loading, sapphire substrates were sonicated for 5~min in acetone followed by isopropanol. The substrates were subsequently annealed \textit{in situ} at 1700~$^\circ$C for 200~s in vacuum to reconstruct the sapphire surface.\cite{Sapphire_annealing} Successful annealing was confirmed by reflection high-energy electron diffraction (RHEED) through observation of the $(\sqrt{31}\times\sqrt{31})\text{R}\pm9^\circ$ sapphire surface reconstruction.

Following annealing, the substrate was cooled to the desired growth temperature and NH$_3$ was introduced into the chamber. The titanium source laser was then ramped to 377~W over approximately 2 min. The titanium growth rate exhibits an Arrhenius-like dependence on laser power;\cite{PowerDependenceOfGrowthRatesEpiray} therefore, the final 50 W of the source ramp was applied rapidly to effectively shutter the molecular flux and provide a reproducible start to film growth. Films were grown for 25 min at a growth rate of approximately 1--3~nm/min. The growth rate varied between experiments due to gradual changes in the titanium source shape and laser mirror reflectivity.

At the completion of growth, the source and substrate lasers were switched off simultaneously. NH$_3$ flow was maintained until the substrate cooled below 200~$^\circ$C, typically within approximately 2~min. Maintaining a nitrogen-rich atmosphere during cooldown is commonly employed for transition-metal nitride growth to suppress nitrogen loss from the film.\cite{Wright2022,NTT_TiN_on_MgO,Smart_TiN_resonator}

\subsection{Characterization}

Structural characterization was performed using X-ray diffraction (XRD), including X-ray reflectivity, symmetric $\theta$-$2\theta$ scans, $\phi$ scans, and $\omega$ rocking curves. The diffractometer was a PANAlytical Empyrean, equipped with a PIXcel\textsuperscript{3D} detector and a hybrid mirror-monochromator incident-beam optics unit that produces a parallel beam of Cu $K$$\alpha_1$ radiation. Symmetric $\theta$-$2\theta$ and $\phi$ scans were collected using a double-axis configuration with a 0.5$^\circ$ divergence slit, while rocking curves were collected using a triple-axis configuration with a two-bounce germanium analyzer crystal to provide higher angular resolution. Atomic force microscopy (AFM) was used to characterize the sapphire surface morphology following NH$_3$ exposure. Scanning transmission electron microscopy (STEM) was performed on selected samples using a Thermo Fisher Spectra 300 aberration-corrected microscope at 300 keV to characterize the TiN/sapphire interface and columnar structure. Room-temperature resistivity and Hall-effect mobility were measured in the van der Pauw geometry using a Lake Shore FastHall station. Temperature-dependent resistivity was measured in a Quantum Design Physical Property Measurement System (PPMS) using aluminum wire bonds. These measurements were used to determine the zero-resistance superconducting critical temperature ($T_\mathrm{c}$) and residual resistivity ratio (RRR), defined as $\rho_{300\mathrm{K}}/\rho_{6\mathrm{K}}$.

\section{Results and Discussion}
\label{sec:results}
\subsection{Optimization of TiN growth}
\label{sec:growth_optimization}

\begin{figure}[b]
    \centering
    \includegraphics[width=\linewidth]{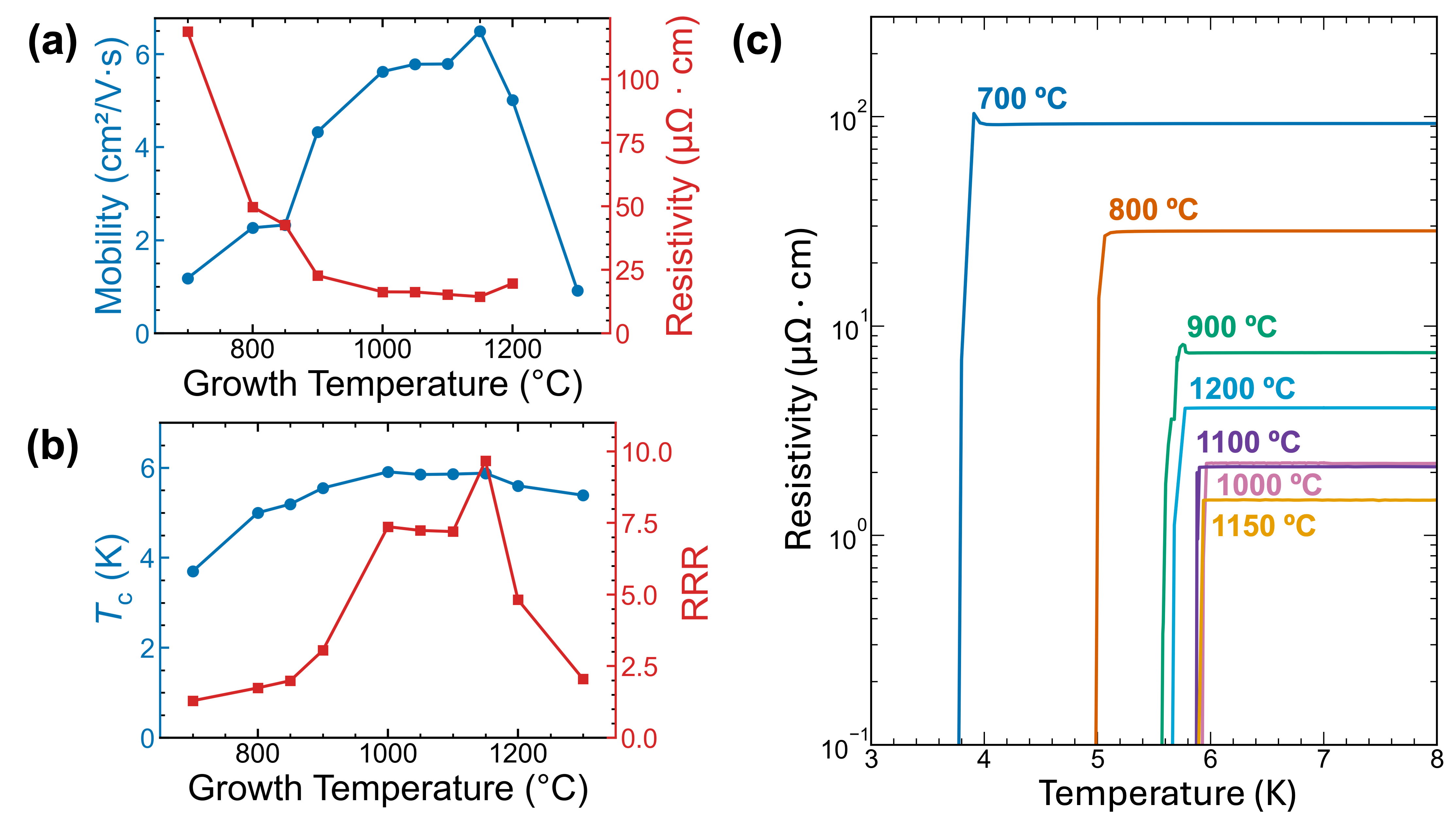}
    \caption{
        Electrical properties of TiN as a function of growth temperature. (a) Room-temperature electron mobility and resistivity. 1300~$^\circ$C resistivity is omitted since the film is too rough to determine thickness using XRR. (b) Superconducting critical temperature ($T_\mathrm{c}$) and residual resistivity ratio (RRR). (c) Superconducting transitions of films grown at different temperatures.
    }
    \label{fig:electrical}
\end{figure}

To establish an optimal growth temperature for device-quality TiN, films were grown between 700 and 1300~$^\circ$C. The lower bound was selected based on the downward trend in conductivity for films grown at lower temperatures, while films grown above 1200~$^\circ$C exhibited multiple out-of-plane orientations. Figure~\ref{fig:electrical}(a) shows that the room-temperature transport properties improve substantially with increasing growth temperature up to 1150~$^\circ$C. The electron mobility increases from 1.2 to 6.5 cm$^2$/V$\cdot$s, while the room temperature resistivity decreases from 119 to 14.4 \textmu$\Omega\cdot$cm between 700 and 1150~$^\circ$C. A similar trend is observed in the cryogenic properties [Fig.\ref{fig:electrical}(b)]. The RRR reaches a maximum of 9.7 at 1150 $^\circ$C, while $T_\mathrm{c}$ plateaus near 5.9 K between 1000 and 1150 $^\circ$C. Above 1150~$^\circ$C, both mobility and RRR decrease sharply, indicating a deterioration in film quality beyond this point. The film grown at 1300~$^\circ$C also exhibited a blue appearance, in contrast to the characteristic gold color of TiN films grown at lower temperatures.\cite{Boltasseva2015Gold} Blue coloration has been attributed to oxygen incorporation\cite{Vaz2003TiNxOy} which may be supplied by the substrate at this elevated temperature.\cite{Kim2024TiODiffusion} 

Figure~\ref{fig:electrical}(c) shows the superconducting transitions of the TiN films. Most films exhibit a sharp decrease in resistivity at $T_\mathrm{c}$, although the films grown at 700 and 900~$^\circ$C show a small increase in resistivity immediately preceding the transition. Together, the transport measurements identify 1150~$^\circ$C as the optimal growth temperature, above which the electrical properties deteriorate rapidly. To determine the structural origin of these temperature dependencies, we next examine the crystallographic orientation and microstructure of the films.

\subsection{Structural Characterization}
\label{sec:Structural Characterization}

\begin{figure*}[t]
    \centering
    \includegraphics[width=\linewidth]{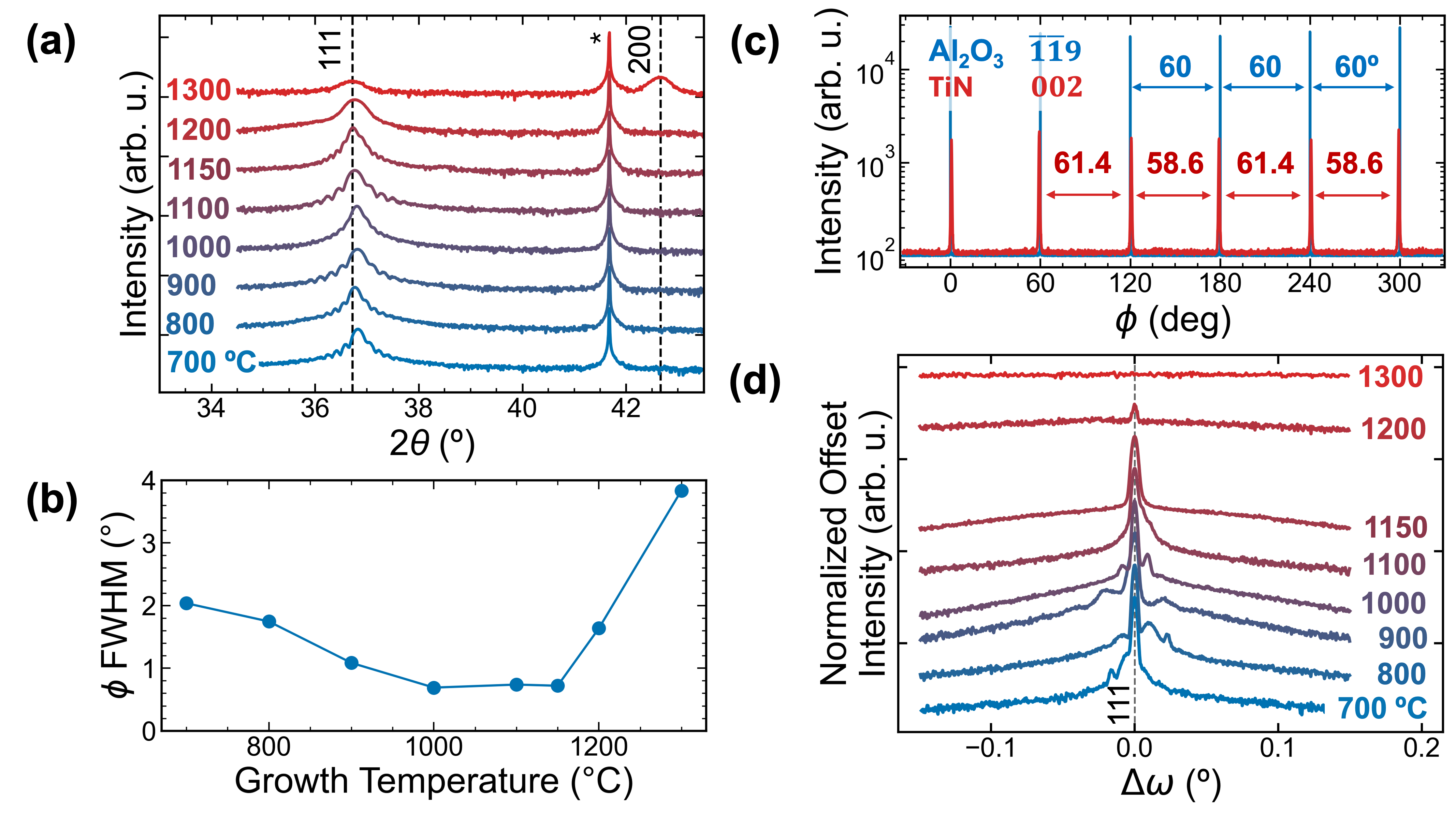}
        \caption
        {
        Structural evolution of TiN with growth temperature. (a) Symmetric $\theta$–$2\theta$ scans. Dashed lines mark the expected TiN 111 and 200 reflections and * denotes the sapphire 006 substrate peak. (b) In-plane mosaic spread determined from asymmetric TiN 002 $\phi$ scans. (c) $\phi$ scan of the film grown at 1150~$^\circ$C showing the epitaxial relationship between TiN and sapphire. (d) TiN 111 $\omega$ rocking curves showing a combination of both narrow and broad components.
        }
    \label{fig:Structure}
\end{figure*}

Structural characterization was performed using XRD and STEM to investigate the origin of the observed trends in electrical properties. Symmetric $\theta$–$2\theta$ scans [Fig.\ref{fig:Structure}(a)] show that films grown from 700 to 1300~$^\circ$C crystallize in the rocksalt $\delta$-TiN phase, with a strong 111 reflection near $2\theta=36.76^\circ$. For bulk TiN ($a=4.235$\si{\angstrom})\cite{NTT_TiN_on_MgO} this reflection is expected at $36.69^\circ$. The position of the TiN 111 reflection shows no systematic dependence on growth temperature and corresponds to an average out-of-plane lattice parameter of approximately 4.23~\si{\angstrom}. The 1300~$^\circ$C film additionally exhibits a TiN 002 reflection near $42.6^\circ$, indicating the emergence of a second out-of-plane orientation at the highest growth temperature.

The films grown between 700 and 1100~$^\circ$C exhibit pronounced Laue oscillations \cite{Laue_1913} surrounding the TiN 111 reflection, signifying coherent diffraction through the film thickness and sufficiently smooth interfaces.\cite{Miller2022Laue} These oscillations are no longer clearly resolved for the 1200~$^\circ$C film, indicating reduced structural coherence. The film nevertheless retains Kiessig fringes,\cite{Kiessig_1931} allowing its thickness to be determined despite an increase in rms surface roughness to 4.2~nm. This is high compared to approximately 1.3~nm roughness for films grown at lower temperatures [Fig.~S1]. At 1300~$^\circ$C, neither Laue oscillations nor Kiessig fringes are clearly resolved, coinciding with a further increase in rms roughness to 6.0~nm and the emergence of the TiN 002 orientation.

The in-plane mosaic spread also varies significantly with growth temperature [Fig.\ref{fig:Structure}(b)]. The full width at half maximum (FWHM) of the $\phi$-scan decreases from $2.0^\circ$ at 700~$^\circ$C to $0.7^\circ$ between 1000 and 1150~$^\circ$C, indicating improved in-plane crystalline alignment. Above 1150~$^\circ$C, the mosaic spread increases sharply, reaching approximately $3.9^\circ$ at 1300~$^\circ$C. This trend closely follows the transport properties, with the lowest in-plane mosaic spread corresponding to the growth-temperature range exhibiting the highest mobility and RRR.

The in-plane epitaxial relationship was examined for the 1150~$^\circ$C film using a $360^\circ$ $\phi$ scan of the off-axis TiN 002 and sapphire $\overline{1}\overline{1}9$ reflections [Fig.\ref{fig:Structure}(c)]. Six TiN 002 peaks are observed despite the threefold symmetry of rocksalt TiN about the [111] growth direction. This signifies two in-plane rotational twin domains commonly observed for rocksalt films grown on sapphire substrates due to double positioning.\cite{CasamentoScN,WrightNbN,TLE_Nitrides,Gao2022TiN} Relative to the expected epitaxial registry, the two twin variants are rotated by approximately $+0.7^\circ$ and $-0.7^\circ$, respectively, causing the alternating TiN peak spacings of approximately $61.4^\circ$ and $58.6^\circ$. A similar alternating rotation is observed for a film grown at 850 $^\circ$C [Fig.~S2], showing the offset is reproducible and not limited to the optimized growth temperature. In contrast, a reference TiN film grown by plasma-assisted MBE (PAMBE) at 600 $^\circ$C substrate temperature exhibits a substantially broader $\phi$-scan FWHM of approximately $3.1^\circ$ and only a small average rotational offset from sapphire of $0.16^\circ$, with both twin variants displaced in the same direction [Fig.~S2]. Because this offset is smaller than the $\phi$-scan step size of $0.2^\circ$, the MBE-grown film exhibits essentially no measurable offset between TiN 002 and sapphire $\overline{1}\overline{1}9$. The high substrate temperatures may contribute to the observed offsets, as these $\phi$-scans are for the highest reported growth temperatures for TiN on sapphire to date. Alternatively, the elevated NH$_3$ pressure used during TLE growth or thermal reconstruction of sapphire's surface may also influence the interfacial structure and resulting epitaxial registry.

Further insight into the crystalline alignment was obtained from $\omega$ rocking curves of the TiN 111 reflection [Fig.\ref{fig:Structure}(d)]. Films grown between 700 and 1200 $^\circ$C exhibit a sharp central component with a Gaussian FWHM of approximately 12–15 arcsec superimposed on a broad diffuse background. A reference measured on a single crystal of silicon using the same configuration exhibited a FWHM of 12 arcsec, indicating that the narrow TiN component approaches the instrumental resolution. The diffuse component flattens with increasing growth temperature and the narrow component disappears completely at 1300~$^\circ$C, corroborating the increased crystalline disorder observed by $\phi$ scans. Similar two-component line shapes have been reported for a variety of heteroepitaxial films exhibiting mosaic disorder.\cite{Miceli_1991,Miceli1995RotationalDisorder,Kortan_1999,Barabash_2001,GaN_rocking,MgN_rocking,Biegalski_2008,Wang_2013,Ruf_2021,Sun2024WO3_rocking,Kalanov2026Ga2O3Rocking} Importantly, Kalanov \textit{et~al.} caution that a resolution-limited central component does not necessarily indicate exceptional crystal quality, as the broad diffuse component reflects the distribution of local crystallographic orientations. The TLE TiN rocking-curve line shapes and broad $\phi$ scans are consistent with the model of rotational disorder proposed by Miceli and Palmstr{\o}m.\cite{Miceli1995RotationalDisorder} They describe locally rotated regions that produce diffuse scattering at short length scales, while the film–substrate interaction limits the magnitude of misorientation at long length scales. This long-range order preserves the structural coherence responsible for the sharp central component.

\begin{figure}[b]
    \centering
    \includegraphics[width=\linewidth]{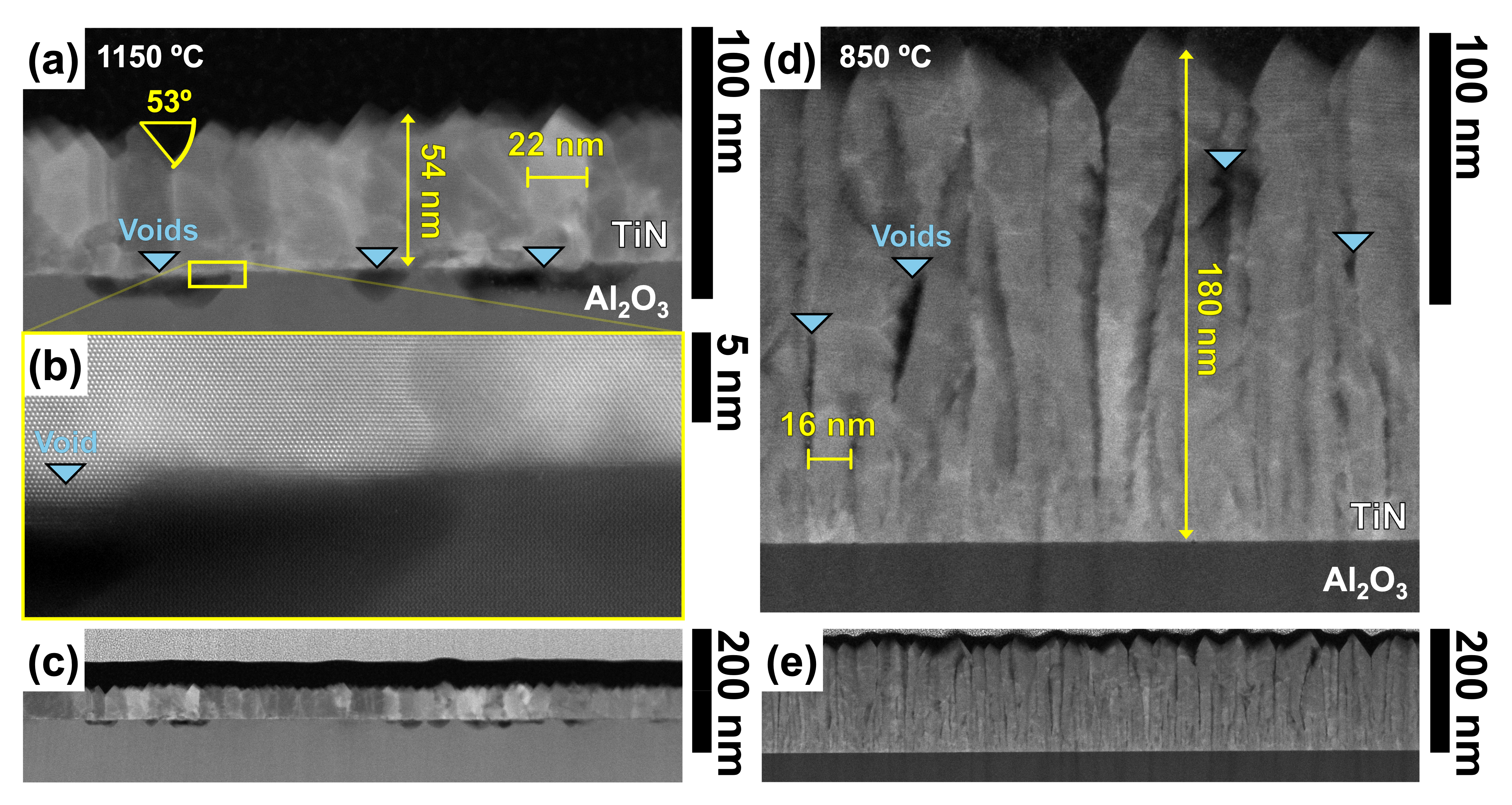}
    \caption{
        Cross-sectional high-angle annular dark-field (HAADF) STEM images of TiN grown on $c$-plane sapphire at (a-c) 1150~$^\circ$C and (d,e) 850~$^\circ$C. The 1150~$^\circ$C film exhibits interfacial voids (highlighted with blue triangles) while the 850~$^\circ$C film maintains a continuous TiN/sapphire interface but has voids nucleating within the film. (b) shows a high-magnification image of a void at the interface. (c,e) show a wide field of view for the films grown at 1150~$^\circ$C and 850~$^\circ$C, respectively.
    }
    \label{fig:STEM}
\end{figure}

Cross-sectional STEM was used to examine the microstructure and TiN/sapphire interface of films grown at 1150 and 850~$^\circ$C [Fig.\ref{fig:STEM}]. Due to variations in titanium flux, the growth rate of the film grown at 850~$^\circ$C was approximately 3 times that of the 1150~$^\circ$C film. The 1150~$^\circ$C film exhibits a faceted columnar microstructure with a column width of approximately 22 nm [Fig.\ref{fig:STEM}(a)]. The column facets form angles approaching $53^\circ$ relative to the (111) interface, consistent with (100) TiN facets. Most importantly, nanoscale voids are distributed along the TiN/sapphire interface, causing local disorder. Because electric fields in superconducting devices are strongly concentrated near the metal–substrate interface,\cite{E_Field_sim,Wang2015Participation} defects in this region can have high participation in dielectric loss. High-resolution imaging of a pitted region reveals the sapphire is viewed along the $m$-axis with the substrate signal disappearing completely in the center of the pit Fig.~\ref{fig:STEM}(b). A wider field of view shows that these pitting defects are pervasive throughout the sample Fig.~\ref{fig:STEM}(c).

In contrast, the film grown at 850~$^\circ$C exhibits a sharp and continuous TiN/sapphire interface across the observed region [Fig.\ref{fig:STEM}(d,e)]. The film retains a columnar microstructure, with a smaller median column width of approximately 16 nm. After 20 nm of continuous growth, voids develop between neighboring TiN columns. This behavior may result from reduced adatom mobility at the lower growth temperature, potentially compounded by the higher growth rate of this sample. Importantly, these voids form within the TiN film rather than at the TiN/sapphire interface. The absence of interfacial voids at 850~$^\circ$C therefore suggests that substrate degradation is strongly temperature dependent. The following section examines the stability of sapphire under the NH$_3$ growth environment as a function of temperature, providing a basis for the two-step TiN growth procedure developed in the final section.

\subsection{Sapphire Nitridation}
\label{sec:nitridation_temperature}

\begin{figure*}[t]
    \centering
    \includegraphics[width=\linewidth]{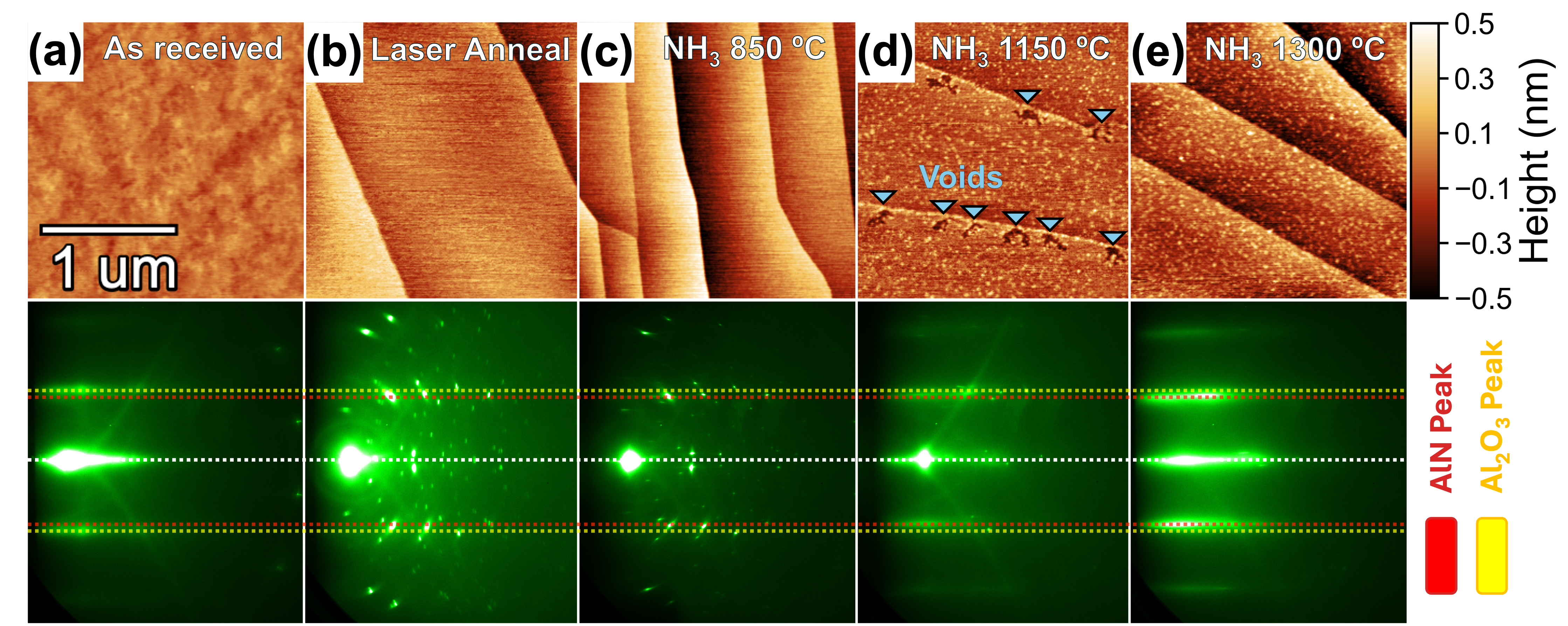}
    \caption{
        Evolution of sapphire surfaces during NH$_3$ exposure. AFM (top row) and corresponding RHEED patterns (bottom row) of (a) as-received sapphire, (b) vacuum-annealed sapphire, and annealed sapphire following $1\times10^{-3}$~Torr NH$_3$ exposure at (c) 850, (d) 1150, and (e) 1300~$^\circ$C. Voids are again highlighted with blue triangles in (d).
    }
    \label{fig:Sapphire_Surfaces}
\end{figure*}

To investigate the origin of the interfacial pitting observed by STEM, vacuum annealed sapphire substrates were exposed to $1\times10^{-3}$~Torr NH$_3$ at different temperatures in the absence of titanium flux. Surface morphology and structural changes were characterized by AFM and RHEED.

Figure~\ref{fig:Sapphire_Surfaces} shows the evolution of the sapphire surface from its as-received state through high-temperature annealing and subsequent exposure to NH$_3$. Vacuum annealing produces smooth, micron-scale terraces [Fig.\ref{fig:Sapphire_Surfaces}(b)]. Following NH$_3$ exposure at 850~$^\circ$C, these terraces are preserved and the reconstructed sapphire RHEED pattern remains visible [Fig.\ref{fig:Sapphire_Surfaces}(c)]. This agrees with the abrupt TiN/sapphire interface observed by STEM at this growth temperature. At 1150 $^\circ$C, nanoscale pits develop preferentially along the sapphire step edges [Fig.\ref{fig:Sapphire_Surfaces}(d)] reminiscent of the interfacial voids observed in STEM. The simultaneous weakening of the sapphire RHEED reconstruction further supports modification of the surface during NH$_3$ exposure. At 1300~$^\circ$C, the original sapphire surface pattern is replaced by streaks consistent with AlN rotated $30^\circ$ in-plane relative to sapphire.\cite{Page2019hBN} These results indicate a strong temperature dependence of the Al$_2$O$_3$–NH$_3$ reaction, progressing from a stable sapphire surface at 850~$^\circ$C to localized pitting at 1150~$^\circ$C and extensive AlN formation at 1300~$^\circ$C. Beyond generating voids, the formation of this reacted surface at high temperature may also contribute to the deterioration of TiN crystalline and electrical properties above 1150~$^\circ$C by disrupting the TiN/sapphire epitaxial interface.

\subsection{Two-Step Growth Recipe}
\label{sec:seed_layer}

\begin{figure*}[t]
    \centering
    \includegraphics[width=\linewidth]{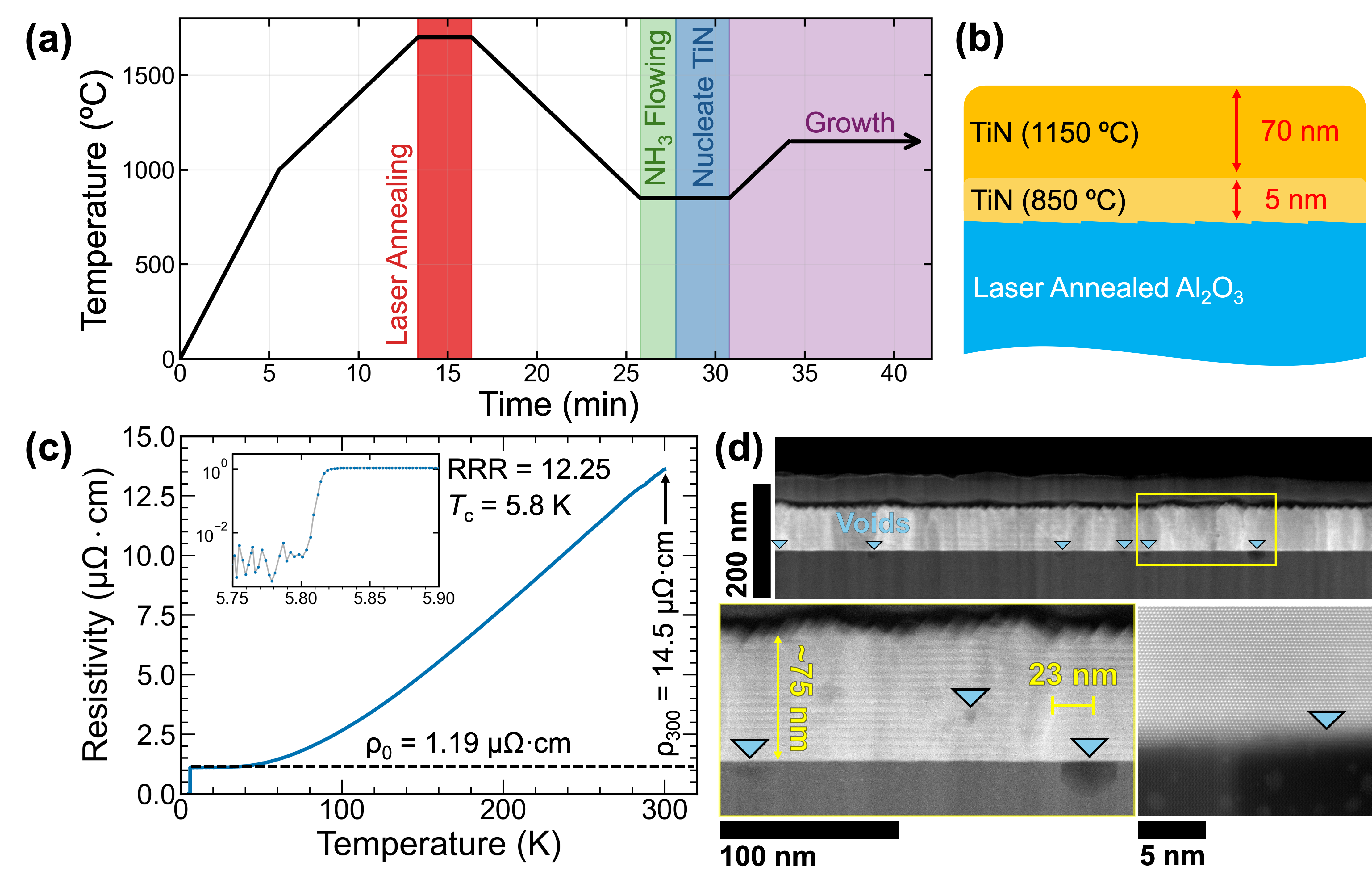}
    \caption{
        (a) Plot of two-step growth recipe. (b) Schematic of the resulting heterostructure. (c) Temperature-dependent resistivity of the two-step grown film, showing $\rho_{300~\text{K}}=13.6~$\textmu$\Omega\cdot$cm, $\rho_0=1.19~$\textmu$\Omega\cdot$cm, and $T_\mathrm{c}=5.8$~K. (d) STEM image of the two-step TiN growth showing improved interface quality.
    }
    \label{fig:Seed}
\end{figure*}

To combine the improved superconducting properties obtained at high growth temperature with a sharper TiN/sapphire interface, a low-temperature TiN nucleation layer was introduced. The resulting growth sequence is shown in Fig.\ref{fig:Seed}a. Following laser annealing, the substrate was cooled to 850 $^\circ$C, where sapphire remains stable under NH$_3$ exposure [Fig.\ref{fig:Sapphire_Surfaces}c]. The NH$_3$ pressure was then stabilized at $1\times10^{-3}$ Torr while the titanium source was ramped to full power over approximately 2~min. TiN was nucleated at 850 $^\circ$C for 3~min, producing a 4–5 nm thick seed layer. AFM measurements show this low-temperature-grown seed layer has a peak-to-valley roughness of approximately 1.5~nm [Fig.~S4], suggesting full coverage of sapphire's surface. The TiN seed thus acts as a barrier between sapphire and the reactive NH$_3$ environment during subsequent high-temperature growth. The substrate temperature was then ramped to the optimized growth temperature of 1150~$^\circ$C at 1.5~$^\circ$C/s, and growth proceeded to the desired film thickness. The resulting structure is illustrated schematically in Fig.~\ref{fig:Seed}b.

To determine how the low-temperature nucleation process affects film properties, temperature-dependent transport measurements were performed [Fig.\ref{fig:Seed}c]. The seeded film exhibits the highest RRR measured in this study, 12.2, while having very similar thickness to its non-seeded counterpart (75 versus 76~nm). This improvement supports diminished defect and interface scattering. The residual resistivity, $\rho_0$, is also the lowest measured in this study at 1.19~\textmu$\Omega\cdot$cm, as determined from a Bloch–Grüneisen fit. The fitted Bloch--Grüneisen temperature of 678 K is comparable to reported Debye temperatures for TiN, which vary between 600--800 K depending on the method and material used for their determination.\cite{Steneteg2013TiN,Kozma2020TiN} The film retains a $T_{\mathrm{c}}$ of 5.8~K, comparable to the highest values obtained without the seed layer. The seeded film also compares favorably with previously reported epitaxial TiN. A recent review by Terai \textit{et al.} reports $\rho_{10\,\mathrm{K}}$ values of 2.6~\textmu$\Omega\cdot$cm for an 80~nm TiN film on sapphire and 1.38~\textmu$\Omega\cdot$cm for a 100~nm film on Si, with corresponding $T_{\mathrm{c}}$ values of 5.63 and 5.67~K, respectively.\cite{Terai2026NitrideQubits} The present film exhibits both a lower low-temperature resistivity and a slightly higher $T_{\mathrm{c}}$ of 5.8~K. Comparable or higher transition temperatures have previously been reported, including $T_{\mathrm{c}}=5.85$~K for TiN on sapphire and values exceeding 6~K for TiN on Si(111).\cite{olson2015growth} Unfortunately, film thicknesses were not reported for these samples, preventing a direct comparison of their low-temperature resistivity.

STEM of the two-step grown film [Fig.\ref{fig:Seed}D] was used to examine the effect of the seed layer on the TiN/sapphire interface. A wide-field image reveals residual interfacial voids. Nevertheless, unlike the clustered pits observed in the unseeded film, these voids are isolated and exhibit reduced contrast relative to the sapphire substrate. Higher-magnification imaging shows that the film microstructure is largely unchanged by the seed layer, retaining similar column widths and a faceted surface morphology. No distinct interface between the low-temperature seed layer and the subsequent high-temperature TiN is discernible by high-resolution STEM within the expected $\sim$5nm seed-layer thickness. This suggests that the seed layer becomes structurally incorporated into the film during subsequent growth at 1150~$^\circ$C. High-resolution imaging also captures a void interrupting an otherwise epitaxial TiN/sapphire interface. In contrast to the more extensively pitted unseeded film, the sapphire $m$-plane remains visible throughout the void, indicating that the substrate has not been completely eroded through the thickness of the lamella. While these observations support an improvement in the structural quality of the interface, some material is still removed from the substrate despite the seed layer. One possible pathway is diffusion of NH$_3$ along the TiN column boundaries to the buried metal-substrate interface, where reactions with the substrate could produce the residual voids.

\section{Conclusions}
\label{sec:conclusions}

Thermal-laser epitaxy enables the growth of high-quality superconducting TiN on sapphire under conditions beyond those typically accessible in conventional molecular-beam epitaxy. An initial temperature series identified 1150~$^\circ$C as the optimal growth temperature, producing films with competitive superconducting and transport properties. Unfortunately, the same high-temperature, ammonia-rich conditions modified the sapphire surface and produced voids at the buried metal-substrate interface. We find that an initial low-temperature (\SI{850}{\degreeCelsius}) TiN seed layer substantially mitigates substrate degradation during subsequent high-temperature deposition while retaining the benefits of high-temperature growth. Utilizing this approach, we achieve a record-low $\rho_{10,\mathrm{K}}$ of 1.19~\textmu$\Omega\cdot$cm, together with a $T_{\mathrm{c}}$ of 5.8~K and an RRR of 12.2. These results indicate that low-temperature nucleation can help circumvent the tradeoff between interface preservation and optimization of the bulk superconducting film, providing a pathway toward sharper interfaces in TLE-grown heterostructures. Future studies will evaluate the impact of these interfacial defects on superconducting resonator performance and compare the microwave loss of high-conductivity TLE-grown TiN with that of MBE-grown and sputtered films.
\section*{Supplementary Material}

See the Supplementary Material for temperature-dependent TiN surface morphologies, additional $\phi$ scans comparing TLE- and MBE-grown TiN, thickness-dependent rocking curves, and further evidence of sapphire's coverage by the low-temperature TiN nucleation layer. 

\begin{acknowledgments}
This work was supported by DOE under Award No.~DE-SC0026541 monitored by Dr. Athena Sefat. Characterization was performed in part at the Cornell Center for Materials Research (CCMR). The authors are grateful to Daniel Tong of the Department of Materials Science and Engineering and Steven Zeltmann of Applied and Engineering Physics at Cornell University for their assistance with STEM lamella preparation for the two-step TiN sample. The authors also acknowledge helpful discussions with Philipp John regarding X-ray rocking-curve analysis at ICMBE 2026.
\end{acknowledgments}

\section*{Author Declarations}

\subsection*{Conflict of Interest}
The author B.D.F. is an employee of epiray GmbH, a company that sells commercial \ce{CO2} laser heater and TLE systems.

\subsection*{Author Contributions}

Hyatt A.: Investigation, formal analysis, visualization, and writing–original draft.

Ithepalli A.: Methodology and writing–review and editing.

Chung C.: Supervision, investigation and writing–review and editing.

Birkh\"olzer Y.: Methodology and writing–review and editing.

Faeth B.: Methodology and resources.

Xing H.: Funding acquisition and supervision.

Muller D.: Supervision, Funding acquisition and resources.

Schlom D.: Resources, funding acquisition, supervision, and writing–review and editing.

Jena D.: Conceptualization, funding acquisition, supervision, and writing–review and editing.

\section*{Data Availability}

The data that support the findings of this study are available from
the corresponding author upon reasonable request.

\bibliography{references}

\end{document}


\title{
Supplementary Material for:
Thermal-Laser Epitaxy of Superconducting TiN on Sapphire:
Growth Optimization and Interface Control
}

\author{Anthony Hyatt}
\affiliation{
Department of Materials Science and Engineering,
Cornell University,
Ithaca, New York 14853, USA
}

\author{Anand Ithepalli}
\affiliation{
Department of Materials Science and Engineering,
Cornell University,
Ithaca, New York 14853, USA
}

\author{Clara Chung}
\affiliation{
Department of Materials Science and Engineering,
Cornell University,
Ithaca, New York 14853, USA
}

\author{Yorick A. Birkh\"olzer}
\affiliation{
Department of Materials Science and Engineering,
Cornell University,
Ithaca, New York 14853, USA
}

\author{Brendan Faeth}
\affiliation{
Department of Materials Science and Engineering,
Cornell University,
Ithaca, New York 14853, USA
}

\author{Huili Grace Xing}
\affiliation{
Department of Materials Science and Engineering,
Cornell University,
Ithaca, New York 14853, USA
}

\author{David A. Muller}
\affiliation{School of Applied and Engineering Physics, Cornell University, Ithaca, NY, USA}
\affiliation{Kavli Institute at Cornell for Nanoscale Science, Ithaca, New York 14853, USA}

\author{Darrell G. Schlom}
\affiliation{
Department of Materials Science and Engineering,
Cornell University,
Ithaca, New York 14853, USA
}
\affiliation{Kavli Institute at Cornell for Nanoscale Science, Ithaca, New York 14853, USA}
\affiliation{Leibniz-Institut f\"ur Kristallz\"uchtung, Max-Born-Str. 2, 12489 Berlin, Germany}

\author{Debdeep Jena}
\email{djena@cornell.edu}
\affiliation{
Department of Materials Science and Engineering,
Cornell University,
Ithaca, New York 14853, USA
}

\maketitle

\section{Temperature-Dependent T\MakeLowercase{i}N Surface Morphology}

Atomic force microscopy (AFM) was used to examine the evolution of TiN surface morphology with growth temperature. Figure~S1 shows representative scans of films spanning the growth-temperature series discussed in the main text. Films grown at substrate temperatures between 700 and 1100~$^\circ$C exhibit comparatively smooth surfaces, with rms roughness values of approximately 1.5~nm or less. The smaller-area scans reveal a granular surface morphology consistent with the columnar microstructure observed by scanning transmission electron microscopy (STEM). A substantial increase in surface roughness occurs above 1100~$^\circ$C, though electronic properties only begin to degrade at 1200~$^\circ$C. The 1200 and 1300~$^\circ$C films have rms roughness of approximately 4.2 and 6.0~nm, respectively. This increase in surface roughness accompanies the loss of pronounced XRD Laue oscillations and the deterioration of the electrical and structural properties discussed in the main text.

\begin{figure}[htbp]
    \centering
    \includegraphics[width=\linewidth]{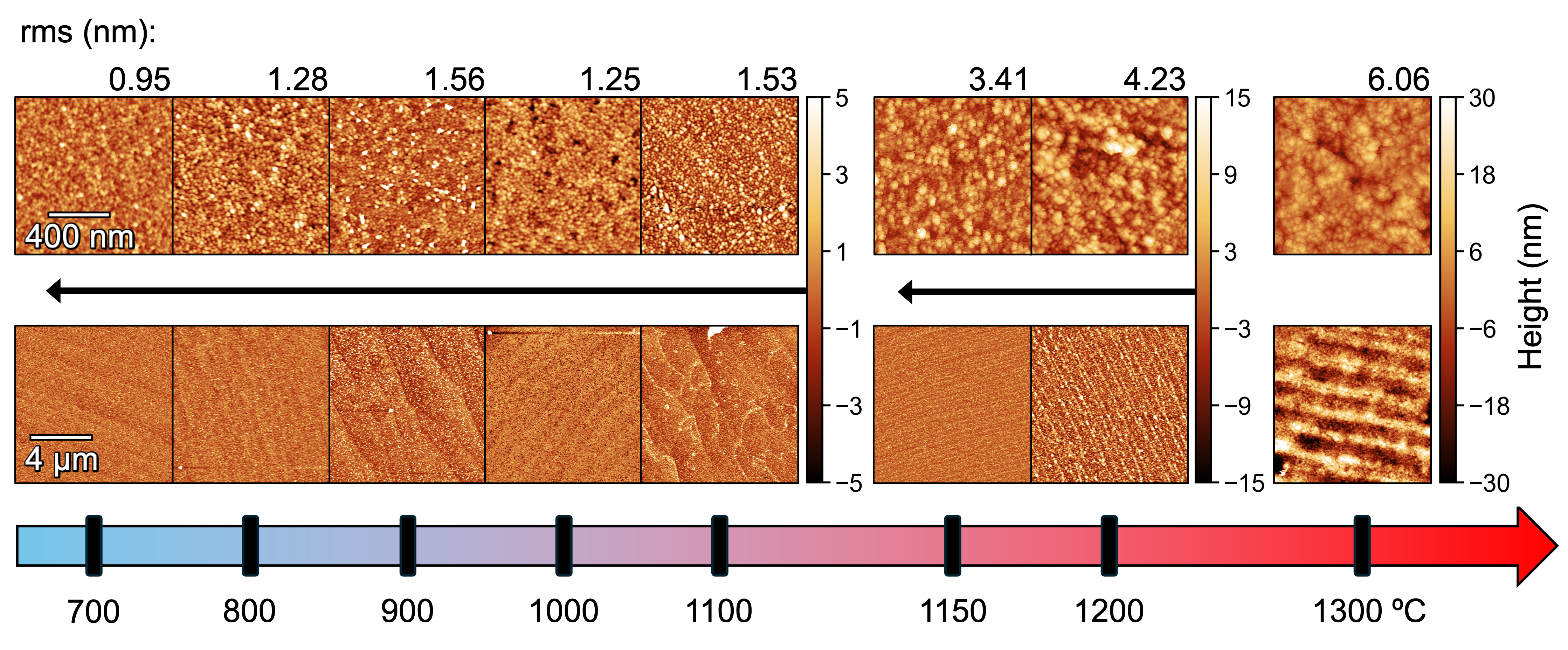}
    \caption{
        \textbf{Surface morphology of TiN as a function of growth temperature.} AFM images of TiN films grown between 700 and 1300~$^\circ$C. The upper row shows smaller-area scans highlighting the local TiN surface morphology, while the lower row shows larger field of view. rms roughness values for the small area scans are denoted above each. Separate height scales are used for the higher-temperature films to accommodate their increased surface roughness. 
    }
    \label{fig:S1}
\end{figure}

\clearpage

\section{Additional T\MakeLowercase{i}N XRD $\phi$ Scans}
Additional $\phi$ scans were collected to determine whether the alternating rotational offset observed for the optimized 1150~$^\circ$C TLE-grown film is reproducible at other growth temperatures. Figure~S2 compares TiN 002 and sapphire $\overline{1}\overline{1}9$ reflections for TLE-grown films and a reference TiN film grown by plasma-assisted molecular beam epitaxy (PAMBE). The twinned epitaxial relationship of TiN on sapphire is also evident in the RHEED pattern of the 1150$^\circ$C film [Fig.~S2(b)], where two sets of diffraction features corresponding to the two in-plane twin variants are observed.

The TLE film grown at 850~$^\circ$C exhibits the same alternating offset of the TiN reflections relative to sapphire observed for the 1150~$^\circ$C film. Thus, the rotation is not unique to the optimized high-temperature growth condition. This sample also has a more centered shoulder on each peak indicating that the film may be mixed between twisted and aligned TiN at intermediate temperatures.

In contrast, the PAMBE-grown reference film exhibits substantially broader TiN reflections, with a $\phi$-scan FWHM of approximately $3.1^\circ$. Its average rotational offset from the expected epitaxial registry is approximately $0.16^\circ$, which is less than the $0.2^\circ$ step size used for the measurement, and the alternating positive and negative rotational offsets observed for the TLE-grown films are absent. These measurements indicate that the alternating rotation is characteristic of the high-temperature TLE-grown samples examined here rather than an unavoidable consequence of TiN growth on sapphire.

\begin{figure}[htbp]
    \centering
    \includegraphics[width=\linewidth]{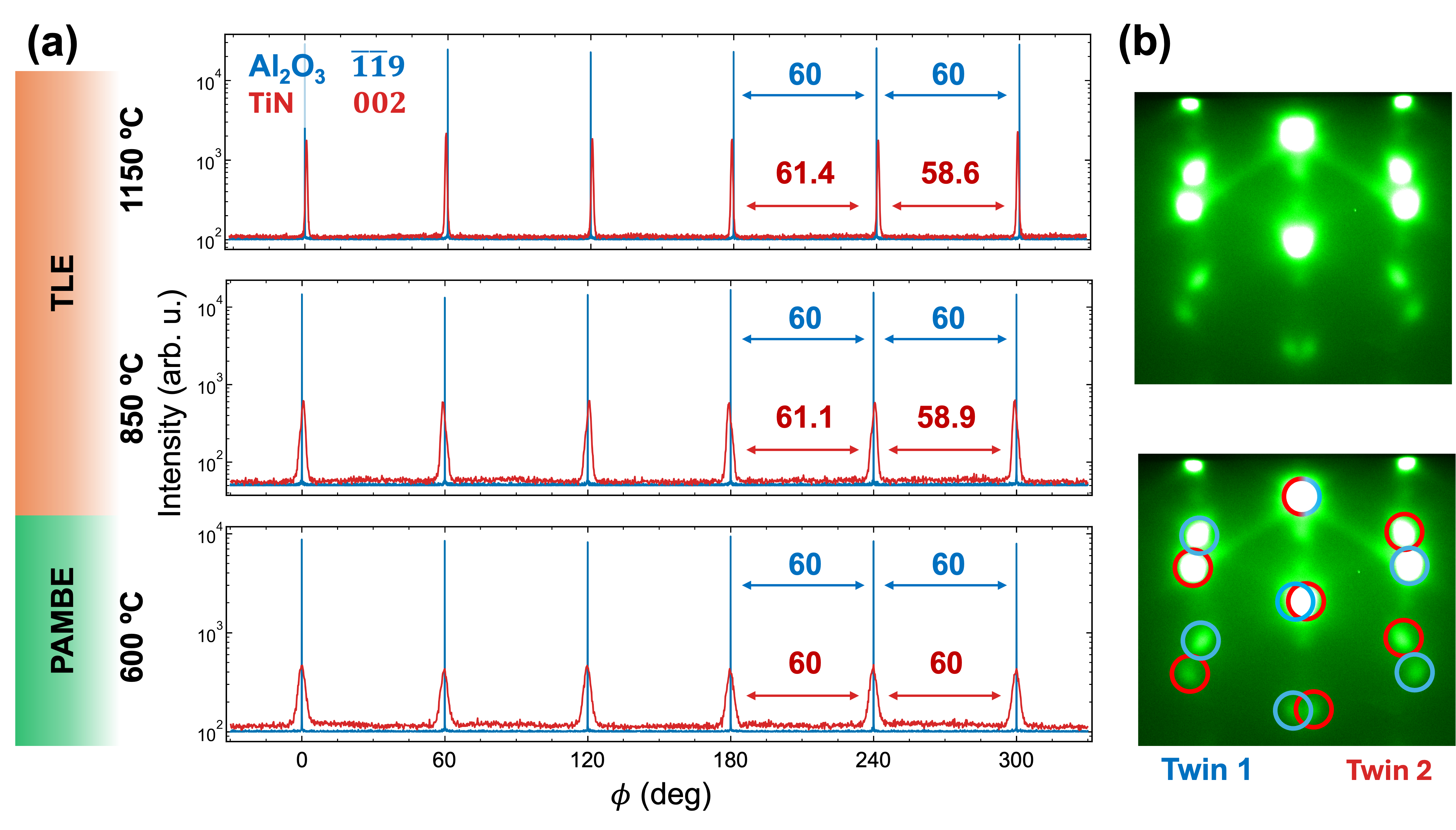}
    \caption{
        \textbf{In-plane epitaxial alignment of TLE- and PAMBE-grown TiN on sapphire.} (a) $\phi$ scans of the off-axis TiN 002 and sapphire $\overline{1}\overline{1}9$ reflections for TiN grown by TLE at 1150 and 850~$^\circ$C and a reference film grown by PAMBE at 600~$^\circ$C. (b) RHEED pattern of the 1150~$^\circ$C TLE-grown film, with the two TiN twin variants identified in the lower image.
    }
    \label{fig:S2}
\end{figure}

\clearpage

\section{Thickness Dependence of T\MakeLowercase{i}N Rocking Curves}

To investigate the origin of the two-component TiN 111 rocking curves described in the main text, films of different thicknesses were grown at 850~$^\circ$C. Figure~S3 shows rocking curves for films with thicknesses ranging from approximately 16 to 180~nm.

The thinnest film exhibits a relatively weak rocking curve with a few distinct angles showing significant intensity. With increasing film thickness, a narrow central peak becomes increasingly prominent while a broad diffuse component develops beneath it. This diffuse component becomes stronger than any specific misorientation angles beyond 58 nm. The diffuse scattering intensity also increases relative to the narrow component by roughly an order of magnitude as the film becomes thicker.

This thickness dependence is consistent with the columnar microstructure observed by STEM.\cite{GaN_rocking,MgN_rocking} Material near the TiN/Al$_2$O$_3$ interface remains highly aligned, producing the narrow component, while increasing column tilt and mosaic disorder during growth contribute progressively to the broad diffuse intensity. The persistence of the narrow component with increasing thickness therefore indicates that the broadening does not originate from a uniform increase in mosaic spread throughout the entire film.

\begin{figure}[htbp]
    \centering
    \includegraphics[width=0.5\linewidth]{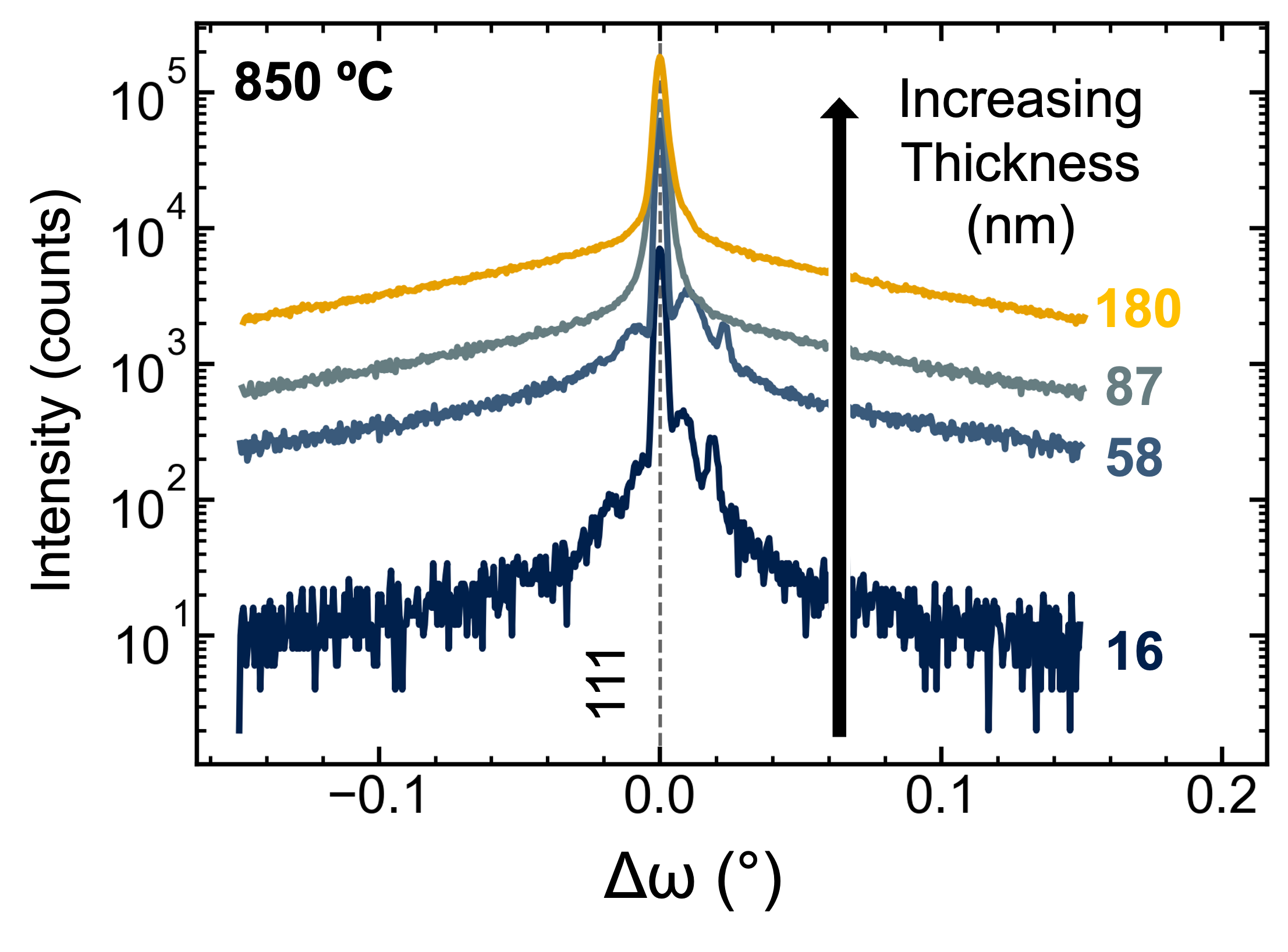}
    \caption{
        \textbf{Thickness dependence of TiN 111 rocking curves.} Rocking curves of TiN films grown at 850~$^\circ$C with thicknesses of approximately 16, 58, 87, and 180~nm.
    }
    \label{fig:S3}
\end{figure}

\clearpage

\section{Coverage of the Low-Temperature T\MakeLowercase{i}N Nucleation Layer}

The continuity of the low-temperature TiN nucleation layer was evaluated using X-ray reflectivity (XRR) and AFM. The seed was grown at 850~$^\circ$C for 3~min using the nucleation procedure described in the main text. XRR of the resulting film is modeled by a TiN layer with an average thickness of approximately 4.2~nm [Fig.~S4(a)].

AFM measurements at two lateral length scales show that the nucleation layer remains conformal with the underlying terraced sapphire morphology [Fig.~S4(b,c)]. The measured peak-to-valley surface variation is approximately 1.5~nm, substantially smaller than the average thickness obtained by XRR. If the full peak-to-valley variation is assigned to variations in TiN thickness, the minimum local film thickness remains greater than zero with a safety margin of 3.4~nm. The combined XRR and AFM measurements therefore support continuous TiN coverage of the sapphire surface following the 3-min low-temperature nucleation step.

Continuous coverage is important because the sapphire surface is then separated from the gas phase before the substrate is heated to 1150~$^\circ$C. The seed layer can consequently suppress direct exposure of sapphire to NH$_3$ during the subsequent high-temperature portion of the growth which is suspected to result in interface voids.

\begin{figure}[htbp]
    \centering
    \includegraphics[width=\linewidth]{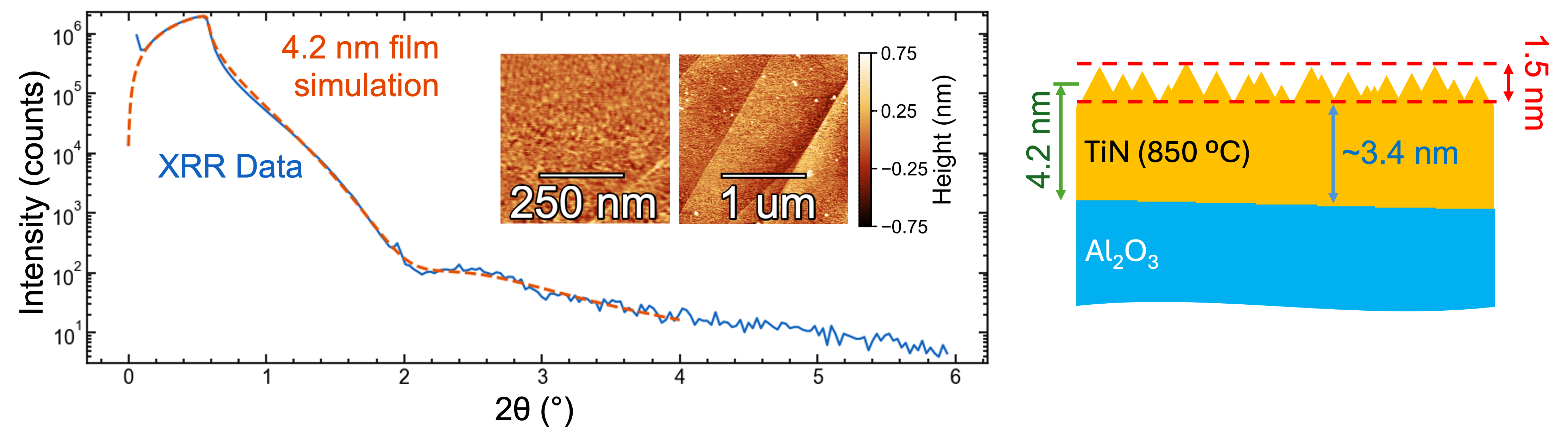}
    \caption{
        \textbf{Coverage of sapphire by the low-temperature TiN nucleation layer.} X-ray reflectivity of a TiN seed layer grown for 3~min at 850~$^\circ$C together with the simulated reflectivity, yielding an average TiN thickness of approximately 4.2~nm. AFM images showing less than 1.5~nm peak to valley roughness are inset. The schematic (right) illustrates the relationship between the average film thickness and the measured surface height variation. 
    }
    \label{fig:S4}
\end{figure}

\bibliography{references}